# A Concept for Autonomous Problem-Solving in Intralogistics Scenarios

Johannes Sigel
Institute of Industrial Automation and Software Engineering
University of Stuttgart
Stuttgart, Germany
johannes.sigel@ias.uni-stuttgart.de

Daniel Dittler
Institute of Industrial Automation and Software Engineering
University of Stuttgart
Stuttgart, Germany
daniel.dittler@ias.uni-stuttgart.de

Nasser Jazdi
Institute of Industrial Automation and Software Engineering
University of Stuttgart
Stuttgart, Germany
nasser.jazdi@ias.uni-stuttgart.de

Michael Weyrich
Institute of Industrial Automation and Software Engineering
University of Stuttgart
Stuttgart, Germany
michael.weyrich@ias.uni-stuttgart.de

*Abstract*—Achieving greater autonomy in automation systems is crucial for handling unforeseen situations effectively. However, this remains challenging due to technological limitations and the complexity of real-world environments. This paper examines the need for increased autonomy, defines the problem, and outlines key enabling technologies. A structured concept is proposed, consisting of three main steps: context enrichment, situation analysis, and generation of solution strategies. By following this approach, automation systems can make more independent decisions, reducing the need for human intervention. Additionally, possible realizations of the concept are discussed, especially the use of Large Language Models. While certain tasks may still require human assistance, the proposed approach significantly enhances the autonomy of automation systems, enabling more adaptive and intelligent problem-solving capabilities.

## I. Introduction

Growing complexity, labor shortages and cost constraints are increasing the pressure to automate industrial systems processes and accelerate the transition to autonomous operations. In this context, industrial systems can be divided into levels from 0 (No Automation) to 5 (Autonomy), similar to the Society of Automotive Engineers (SAE) levels for autonomous driving [1]. Higher levels of autonomy promise significant benefits, including increased efficiency, reduced operational costs, enhanced safety, and the ability to operate in hazardous environments without human intervention [2]. By leveraging advanced sensing, data-driven decision-making, and self-optimization capabilities, autonomous systems could transform industrial operations and improve overall system resilience [3].

However, despite these potential benefits, fully autonomous industrial systems remain largely unrealized [4]. Key challenges include the integration of heterogeneous systems, ensuring robustness and reliability under dynamic conditions, addressing safety concerns, and managing the complexities of human-machine collaboration. Additionally, existing automation solutions often lack the adaptability required to handle unforeseen events, limiting their practical applicability in real-world scenarios [5].

This paper therefore addresses the question:

*How can automation systems correctly assess situations in dynamic, unknown environments and find solutions to problems that occurred in their environment on their own?*

The rest of this paper is structured as follows: Section II introduces the problem scenario and examines the challenges. The current state of the art is given in Section III. Section IV describes the general overview of the concept, possibilities for the realization are discussed in Section V. Finally, Section VI gives a short summary and outlook.

## II. Initial Situation and Problem Formulation

This section provides a short background on the context of the work and the problem, as well as the area of industrial automation where the concept can be applied. An interesting use case can be found in the intra logistics sector, for which a demonstration model is available at our institute. In a manufacturing or storage facility, various autonomous transport vehicles operate within a network. Each is an automation system on its own, collectively forming a "system of systems (SoS)" [6]. These vehicles move goods between warehouses and production areas, with different vehicles suited to different sizes and shapes. One such unmanned vehicle is on its way to its destination when it encounters an obstacle, such as a pallet (Fig. 1). Unable to navigate around or move the pallet, the vehicle must stop and wait for a human worker to clear the path.

In order for the system of transport vehicles to operate autonomously, it must develop a plan for the independent removal of obstacles. In the presented scenario, an unmanned transport vehicle capable of transporting pallets can be instructed to move the pallet (see Fig. 2). Nonetheless, in the case that this type of task has not been explicitly modeled within the system, the system itself must ascertain the available resources and determine the necessary actions to resolve the issue that has emerged. To achieve this objective, the system must not only evaluate various pieces of information but also understand the context of the situation and determine how the available resources can be used.

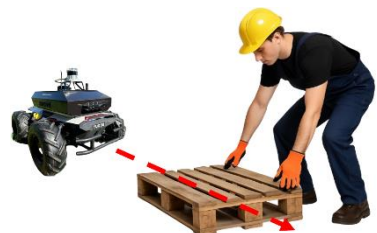

Fig. 1. State of the art: wait for human interaction

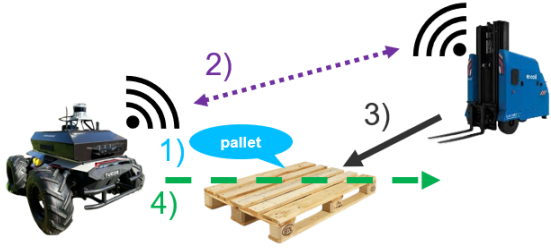

Fig. 2. The system solves the problem itself

In conclusion, autonomous systems must perform the following steps:
- Collect sensor data and enrich it with *context*.
- Detect the reason why the goal cannot be achieved and analyze possible further problems through *anomaly prediction*.
- Generate possible solution strategies to solve the problem and to evaluate which one is the best, for which *situation awareness* is necessary.

The concept presented in this paper integrates these approaches to enhance the autonomy in automation systems. Therefore, the state of the art in these fields is presented in the next chapter.

## III. STATE OF THE ART

As seen in the preceding chapter, autonomous systems capable of solving unforeseen problems require a combination of multiple technologies to perceive, analyze, and adapt to dynamic environments. The concept presented in this paper builds upon and tries to improve various existing approaches to enhance autonomy in automation systems. To provide the necessary background, this chapter outlines the state of the art and the remaining challenges for the key technologies involved.

### A. Context in industrial automation

To perform a more effective analysis and particularly to relate multimodal data to one another or similar data from various viewpoints, it is essential to augment the data with contextual elements.

There is no agreement on a universal definition of context, as it differs in range and is predominantly application-focused [7]. From a linguistic perspective, context can be described as the implicit information available that aids in interpreting and understanding interactions and communication messages [8]. In [9] Sahlab et al. defined a "context tier model (CMT)" to show four different abstraction levels of system context. In the scope of this paper, the levels C3 "System of Systems context" and C4 "External System Context" are relevant, as context is needed as supplementary information regarding objects or other systems to clarify how the system might or might not engage with these objects or other systems. This supplementary information may encompass metadata, relationships, and attributes that deepen the understanding of the object's function within the system. To improve the contextual annotations of the gathered data, the different systems should exchange their context knowledge about their surroundings. But a brief look into the literature suggests that there is still a research gap in this area. For instance, Müller et al. used the CMT approach to consolidate and connect diverse internal and external data for context-enhanced modeling in intelligent Digital Twins (iDTs), but there is no data exchange between different iDTs [10]. In [11], Vu et al. conducted a literature research with the result, that most of the analyzed studies only try to acquire context information from raw data. To improve the response generation in dialogue systems, Dong et al. designed a method to exchange graph-based knowledge between the user model and an LLM [12]. While this concept is tailored for dialogue systems, its core principles might inspire future adaptations for industrial automation. However, the current state of the art remains unsatisfactory in terms of the exchange of information with the aim of improving the context annotations of gathered sensor data.

### B. Anomaly prediction

To plan ahead and not only react to events, it is crucial to not only detect but also predict situations that might occur. This procedure entails recognizing patterns or actions that stray from the usual and is commonly referred to as anomaly detection, or rather in this case, anomaly prediction.

Anomalies pose a significant challenge as they are unknown, irregular, rare, and vary widely in type [13]. This complexity has driven extensive research efforts to develop effective methods for detecting and predicting anomalies. Many modern approaches rely on machine learning techniques. An overview of the prevailing methods in this domain has been delineated by Yang et al. in [14]. Here, traditional methods are described, but also emerging techniques based on LLMs are presented. The key advantages for leveraging LLMs for anomaly detection are semantic understanding, contextual awareness, and adaptability, enabling a deeper and more nuanced detection of anomalies. Indeed, contextual awareness remains challenging in the prediction of anomalies. As asserted in [15], the ability to adapt to changing contexts is especially important when dealing with shifting data patterns over time. Therefore, incorporating additional context can enhance the robustness and accuracy of anomaly detection and prediction systems by enabling them to better interpret and respond to dynamic data environments.

### C. Situation awareness

In recent years, the concept of situation awareness has gained significant attention in the field of industrial automation, as it plays a crucial role in enhancing safety, efficiency, and decision-making processes within complex systems.

The process of understanding the situation in which the system finds itself is called situation awareness. This concept is based on situation awareness by humans, as described by Endsley [16], who defined it as the perception of elements in the environment, comprehension of their meaning, and projection of their status into the future. Müller et al. offer a formalized definition of situation awareness as a function that correlates a collection of measurements and the present situation to the context and upcoming situation [17]. The concept of situation awareness is already used in various industrial automation applications, for example, for the improvement of human-robot collaborations [17], for monitoring, reconfiguring and maintaining of production lines [18] or to evaluate situative risks in trajectory planning [19]. A gap in research persists regarding the prediction of future scenarios and the efficient integration of multimodal data [20]. Izquierdo et al. state that the evaluation of a model's accuracy is also context-dependent, which complicates the standardization of assessment methods across different applications [18]. Addressing these challenges is crucial for advancing situation awareness technologies and ensuring their reliability in dynamic industrial environments.

## D. Conclusion

In conclusion, this chapter explored context enrichment, anomaly prediction, and situation awareness, highlighting key challenges in each area. The analysis shows that enhancing systems with richer contextual information is crucial for understanding system dynamics and enabling better decision-making. To address these challenges, the next chapter introduces a concept that uses a knowledge base and enables communication between Digital Twins, aiming to improve contextual awareness and system responsiveness.

## IV. CONCEPT FOR AUTONOMOUS PROBLEM SOLVING

To ensure the safe and efficient operation of a system, it is preferable to use rule-based algorithms, and in most cases, the system can operate quite well according to its predefined rules. However, if an exceptional situation arises due to a failure, anomaly, or other problem, the implemented rules may not be able to handle the situation.

Based on the problem definition from chapter II and the challenges described in chapter III, the following requirements can be placed on a possible concept for autonomous problem solving:

- The system must be able to recognize unknown situations, react at runtime, adapt and evaluate alternative solutions to ensure situational awareness in new scenarios.
- The system must recognize objects and their relationships to each other, update its knowledge base and adapt accordingly.
- The system must comply with safety regulations and react in real time, especially to critical actions.

The concept proposed in this paper consists of a Digital Twin (DT) according to the definition in [21], as this kind of DT is widely used in industrial automation [22]. With this DT, the following steps must be taken after the detection of a problem that forces the system out of its rule-based states:

- First, various data must be collected, processed (i.e., object recognition in camera images), and enriched with context to describe the situation. Integrating data from additional systems may be helpful.
- Second, information from the first step is used to understand the situation, prioritize the information, and predict future events.
- Third, possible solution strategies are to be generated and evaluated with the help of the information from the previous steps. The most reasonable strategy is then executed.

These steps are carried out in parallel in every subsystem. Through communication between the different systems, they exchange relevant data for their current situations. This concept is shown in Fig. 3 and is described in more detail below.

## A. Context Enrichment

To assess the current situation in which the system is operating, multimodal sensor data from the environment is continuously gathered to provide a comprehensive overview. Raw sensor inputs, such as camera images or LiDAR scans, are preprocessed using established algorithms, for instance, object detection methods classify and localize obstacles within visual data. The resulting structured data is then categorized into static elements, dynamic objects, and critical components requiring heightened attention for safety. Furthermore, the data is enriched with contextual knowledge from the knowledge base, allowing the system to infer, for example, that a pallet is transportable by a forklift or that a human may respond to an auditory or visual signal. With this additional external system context, the DT of the system and its environment can be used for predictive modeling and simulation-based decision-making. If the system's underlying knowledge base is not sufficient, other systems can send relevant data to be added, ensuring an up-to-date knowledge base in a changing environment.

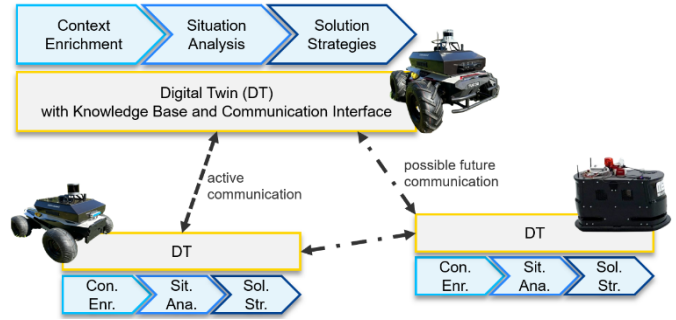

Fig. 3. The concept is executed in parallel in each subsystem

## B. Situation analysis

The generated situational description from the first step is analyzed with respect to the system's goal, ensuring that only relevant information is considered for decision-making. To distinguish between essential and negligible data, the system must first assess what it aims to achieve and identify the factors currently obstructing this objective. This requires pinpointing the root cause of any detected problem, such as an obstructed pathway or a conflicting task allocation. The objective is to detect both the current disruptions and possible future anomalies. The latter might indicate future actions by the systems. By understanding these underlying issues, the system can derive the necessary actions to resolve the initial problem efficiently, enabling adaptive and goal-oriented behavior.

## C. Solution Strategies

Based on the DT of the system and its environment, the defined goal, the identified causes preventing it from being achieved, and possible future anomalies, different strategies for solving the problem are generated and systematically evaluated. To accomplish this, the system must have precise knowledge of its available resources, including e.g. vehicles, robotic arms, and storage capacities, as well as executable actions, such as navigating around an obstacle, rerouting transport tasks, or rescheduling workflows. The DT uses this information to simulate different possible action sequences and their outcomes, allowing the system to explore alternative solutions before making a decision. Furthermore, communication between the DT and other subsystems can facilitate potential collaborative efforts among the various systems. The selection of the most suitable strategy is based on predefined optimization criteria, which may include safety considerations, minimizing energy consumption, reducing transport time, or balancing workload efficiency. The optimal strategy is then translated into executable commands and carried out by the real hardware.

## V. REALIZATION

For the evaluation of the presented concept a prototype will be developed in simulation and with real hardware. The following discourse offers a concise overview of possible technologies, with a particular focus on Generative Artificial Intelligence (GenAI).

For the system to fulfil the requirements of Chapter IV, which relate to the context and relationship of objects in unknown environments, the underlying knowledge base of the DT of a subsystem must be adaptable and expandable. Two approaches are employed for this purpose: With the help of the broad knowledge of Large Language Models (LLMs) or Visual Language Models (VLMs), the context of objects that were not originally available in the knowledge base of the DT can be captured. In addition, the different systems - or rather their DTs - should be able to exchange information about their knowledge of the external system context, as well as for the System of Systems context. Achieving this objective between disparate subsystems necessitates the implementation of a general interface, a task that could be facilitated through LLM-assisted interface generation or translation.

To analyze the situation using anomaly prediction, the use of LLMs can also be employed, as previously described in Chapter III. This approach enables the optimal use of the stored external system context of the knowledge base. Given the presence of graphs in the knowledge base, which are based on ontologies, the utilization of Large Graph Models (LGMs) [23] may prove advantageous. Furthermore, the reasoning capability of these LLMs can be employed to generate potential combinations of individual system actions as solution strategies, similar to the approach outlined by Xia et al. in [24], who applied LLM agents to generate coordinated system actions.

However, the disadvantage of using GenAI is obvious: Hallucinations are still not completely ruled out [25] and therefore the results of these algorithms have to be checked for actions that are critical to safety.

## VI. CONCLUSION AND OUTLOOK

Autonomous problem-solving is essential for increasing the adaptability and efficiency of automation systems, especially in unforeseen situations. However, achieving a higher degree of autonomy remains a challenge due to limitations in current technology and the complexity of real-world environments. The key findings of this paper are the following:
- The general problem with highly automated systems is situations that are not considered in the design phase of the system.
- The key technologies for achieving high autonomy are context enrichment, anomaly prediction and situation awareness.
- The proposed concept uses digital twins with a structured approach consisting of three essential steps: context enrichment, situation analysis, and generation of solution strategies.
- Possible technologies for the realization of the proposed concept are discussed.

Following a successful trial, the concept will also be used for real Unmanned Ground Vehicles (UGVs) at the University of Stuttgart (see Fig. 3). These UGVs have powerful computer resources on board and they can communicate with a local server. Different scenarios can be implemented and evaluated as several UGV variants are available.

Furthermore, a catalogue of intralogistics scenarios is being planned. While some situations may still require human intervention (if the whole system is lacking the necessary resources) the automation system should at minimum possess the capacity to detect its operational boundaries and proactively request targeted human support. Despite remaining challenges, the proposed approach represents a significant step toward more autonomous and intelligent automation systems.